\documentstyle[twoside,fleqn,espcrc2,epsf]{article}

\newcommand{\AmS}{{\protect\the\textfont2
  A\kern-.1667em\lower.5ex\hbox{M}\kern-.125emS}}

\newcommand{\ix}[1]{$#1$}

\newcommand{\benum}{\begin{enumerate}}
\newcommand{\eenum}{\end{enumerate}}
\newcommand{\bit}{\begin{itemize}}
\newcommand{\eit}{\end{itemize}}
\newcommand{\be}{\begin{equation}}
\newcommand{\ee}{\end{equation}}
\newcommand{\bdm}{\begin{displaymath}}
\newcommand{\edm}{\end{displaymath}}
\newcommand{\bqq}{\begin{eqnarray}}
\newcommand{\eqq}{\end{eqnarray}}
\newcommand{\barr}{\begin{array}}
\newcommand{\earr}{\end{array}}

\newcommand{\resone}
{
$\metapr$=$0.960(87)^{+0.036}_{-0.286}$ GeV
}
\newcommand{\etapr}{{\eta^{\prime}}}

\newcommand{\Nf}{N_{\rm{f}}}

\newcommand{\mev}{\,\rm{Me\kern-0.1em V}}
\newcommand{\MeV}{\,\rm{Me\kern-0.1em V}}
\newcommand{\gev}{\,\rm{Ge\kern-0.1em V}}
\newcommand{\GeV}{\,\rm{Ge\kern-0.1em V}}

\newcommand{\mpi}{m_{\pi}}

\newcommand{\Gpi}{G_{\pi}}

\newcommand{\Getapr}{G_{\etapr}}

\newcommand{\Gonlp}{G_{\rm 1}}
\newcommand{\Gtwlp}{G_{\rm 2}}

\newcommand{\metapr}{m_{\eta^{\prime}}}

\newcommand{\Nnoise}{N_{\rm noise}}

\newcommand{\Qtopo}{Q_{\rm topo}}

\newcommand{\alatt}{a}

\newcommand{\Orda}{{\rm O}(\alatt)}

\newcommand{\hlf}{\frac{1}{2}}

\newcommand{\csw}{c_{\rm sw}}

\newcommand{\ident}{\rlap{1}\kern0.2em\rm{l}}
\newcommand{\identn}[1]{\rlap{1}\kern0.2em\rm{l}_{#1}}
\newcommand{\identalbe}{\rlap{1}\kern0.2em\rm{l}_{\alpha\beta}}
\newcommand{\identcd}{\rlap{1}\kern0.2em\rm{l}_{cd}}
\newcommand{\identef}{\rlap{1}\kern0.2em\rm{l}_{cd}}

\newcommand{\Reals}{\rlap{1}\kern0.05em\rm{R}}

\newcommand{\pslash}{\rlap{$p$}\kern0.2em\rm{/}}
\newcommand{\qslash}{\rlap{$q$}\kern0.2em\rm{/}}
\newcommand{\Aslash}{\rlap{$A$}\kern0.2em\rm{/}}
\newcommand{\Aslashcd}{\rlap{$A$}\kern0.2em\rm{/}_{cd}}
\newcommand{\Dslash}{\rlap{$D$}\kern0.2em\rm{/}}
\newcommand{\Dslashcd}{\rlap{$D$}\kern0.2em\rm{/}_{cd}}
\newcommand{\Dslashcdalbe}
{\rlap{$D$}\kern0.2em\rm{/}_{cd\alpha\beta}}
\newcommand{\Dslashalbecd}
{\rlap{$D$}\kern0.2em\rm{/}_{\alpha\beta cd}}
\newcommand{\partslash}{\rlap{$\partial$}\kern0.2em\rm{/}}

\renewcommand{\GeV}{\mbox{GeV}}
\renewcommand{\MeV}{\mbox{MeV}}

\title{
Results for the \ix{\etapr} Mass from Two-Flavor Lattice QCD  
}
 
\author{
CP-PACS Collaboration:
   V.~I.~Lesk\rlap,\address{
              Center for Computational Physics,
              University of Tsukuba, Tsukuba, Ibaraki 305-8577, Japan}
    S.~Aoki\rlap,\address{
              Institute of Physics,
              University of Tsukuba, Tsukuba, Ibaraki 305-8571, Japan}  
    R.~Burkhalter\rlap,$^{\rm a}$ 
    M.~Fukugita\rlap,\address{
              Institute for Cosmic Ray Research,
              University of Tokyo, Kashiwa 277-8582, Japan}
    K.\-I.Ishikawa\rlap,$^{\rm a,b}$
    N.~Ishizuka\rlap,$^{\rm a,b}$
    Y.~Iwasaki\rlap,$^{\rm a}$ 
    K.~Kanaya\rlap,$^{\rm a}$ 
    T.~Kaneko\rlap,$^{\rm d}$ 
    Y.~Kuramashi\rlap,$^{\rm d}$
    M.~Okawa\rlap,\address{
              Department of Physics, Hiroshima University,
              Higashi-Hiroshima 739-8526, Japan} 
    Y.~Taniguchi\rlap,$^{\rm b}$
    A.~Ukawa\rlap,$^{\rm a,b}$ 
    T.~Umeda\rlap,$^{\rm a}$
and T.~Yoshi\'e$^{\rm a,b}$}

\begin{document}

\begin{abstract}
We present results for the mass of the $\eta^\prime$ meson for two-flavor
lattice QCD in the continuum limit, calculated on the CP-PACS computer,
using an RG-improved gauge action and clover fermion
action with tadpole-improved $\csw$.
Measurements are made at three couplings
corresponding to $a\approx 0.22, 0.16, 0.11$~fm
for four quark masses corresponding to
$m_\pi/m_\rho\approx 0.8, 0.75, 0.7, 0.6$.
The two-loop diagrams are evaluated using a noisy source method.
Quark smearing for both one- and two- loop diagrams 
is successfully applied to obtain ground state signals in the $\etapr$ channel.
We obtain \resone in the continuum limit,
where the second error represents the systematic uncertainty coming
from varying the functional form for chiral and continuum extrapolations.
\end{abstract}

\maketitle
\section{Introduction}
An accurate lattice calculation of 
\ix{\metapr} would distinguish between 
QCD and quark models, and provide information relevant to the U(1) problem.
Those attempts
carried out so far \cite{
sesam}, however,
have employed the quenched approximation or used 
only one value of lattice spacing \ix{\alatt} for full QCD.
In this work we make a step forward by using \ix{\Nf=2} full QCD
at several $a$, and evaluating \ix{\metapr} in
the continuum limit.

A technical difficulty which is important to overcome
is the poor signal-to-noise ratio of the $\eta^\prime$
propagator \ix{\Getapr(t)}. 
Its error increases quickly with the time separation \ix{t}
so that an effective mass plateau is not in general
observed when a point quark source is used.
One sometimes fits \ix{\metapr} from the ratio \ix{\Getapr(t)}/
\ix{\Gpi(t)}$=1-A{\rm exp}((\metapr-\mpi)t)$, hoping for
cancellation of effects of excited states.
However there is no rationale behind this expectation.
We improve upon this by employing a smearing method which
enables us to obtain ground state signals in the $\eta^\prime$ 
channel for all simulation parameters.


After a description of our method in Sec.~\ref{sec:method},
we evaluate in Sec.~\ref{sec:result} \ix{\metapr} including systematic error
from chiral and continuum extrapolations.
A correlation between topology and \ix{\metapr} which we have
observed is presented in Sec.~\ref{sec:topology}.

\section{Calculational Method}
\label{sec:method}

We use \ix{\Nf=2} gauge configurations previously
generated for our study of the spectrum of ordinary 
hadrons~\cite{cppacsspectrum},
with an RG-improved gluon action 
and clover quark action with tadpole-improved \ix{\csw}.  
We use three couplings \ix{\beta=1.8}, \ix{1.95} and \ix{2.1} corresponding to 
$a\approx 0.22$, 0.16, 0.11 fm, and four hopping parameters \ix{\kappa} 
matched to correspond to $m_\pi/m_\rho\approx 0.8$, 0.75, 0.7, 0.6. 
The box size $L_{\rm s}a\approx 2.5$~fm has also been
matched between \ix{\beta}.
Valence and sea quark masses are identical.
We take $(\bar u u + \bar d d)/{\sqrt 2}$ as an
operator for the \ix{\etapr}. 

\ix{\Getapr(t) \equiv \Gonlp(t) - \Gtwlp(t)} is
calculated on 400--800 stored configurations for each
$(\beta, \kappa)$ pair, where 
\ix{\Gonlp(t)\equiv \Gpi(t)} (\ix{\Gtwlp(t)}) is the
contribution of the one- (two-) loop diagram.
We use an exponential-like smearing function
given in Ref.~\cite{cppacsspectrum} for both \ix{\Gonlp}
and \ix{\Gtwlp}. One or two quark sources are smeared, while 
sink is always local.
Errors are estimated by the jackknife method using bins of
50 trajectories.
The lattice scale is fixed from $m_\rho$.

%



\begin{figure}[t]
\begin{center}
\centerline{ \epsfxsize=7.0cm 
\epsfbox{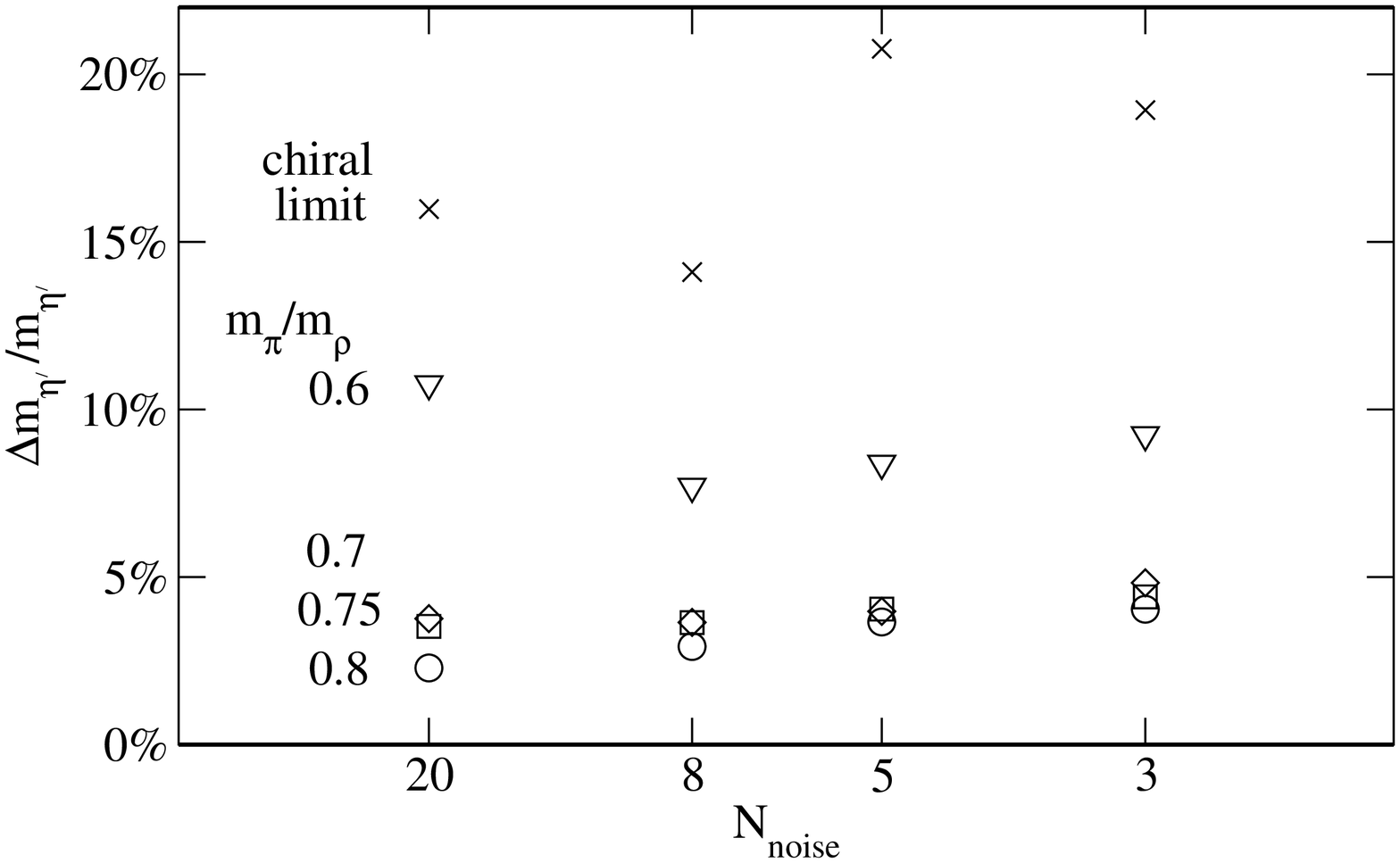} }
\vspace{-12mm}
\caption{Relative error of $\metapr$ versus \ix{\Nnoise}. x-scale is
proportional to \ix{1/\sqrt{\Nnoise}}.}
\label{fig:noisedependence}
\vspace{-14mm}
\end{center}
\end{figure}

\begin{figure}[t]
\begin{center}
\vspace{-5mm}
\centerline{ \epsfxsize=7.0cm \epsfbox{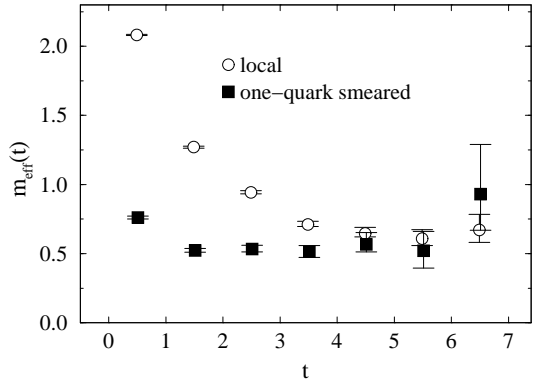}}
\vspace{-12mm}
\caption{Comparison of \ix{\etapr} effective masses
obtained from local and smeared sources at $\beta=$ 2.1 and 
$m_\pi/m_\rho\approx $ 0.7. \ix{m_{\rm eff}(t)} comes
from \ix{G(t-\hlf)} and \ix{G(t+\hlf)}.}

\label{fig:effectivemass}
\vspace{-14mm}
\end{center}
\end{figure}

After fixing configurations to the Coulomb gauge, we
estimate \ix{\Gtwlp(t)} using a noisy
source method with \ix{U(1)} random fields. 
In order to enhance signals, inversions for the quark
propagator are carried out independently for each
(color,spin) index. 
Fig.~\ref{fig:noisedependence} obtained for the coarsest
lattice at $\beta=1.8$
shows that increasing the number of noisy
samples from $\Nnoise=3$ to 20 gives no substantial gain
in precision in \ix{\metapr}.
We thus choose \ix{\Nnoise=3} for finer lattices at
$\beta=1.95$ and 2.1. At \ix{\beta=1.8}, we use results
with \ix{\Nnoise=8}, giving the smallest error in the chiral limit.




\begin{figure}[t]
\begin{center}
\centerline{ \epsfxsize=7.0cm 
\epsfbox{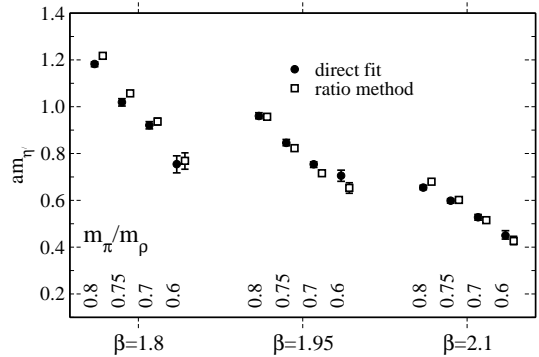} }
\vspace{-8mm}
\caption{Comparison of \ix{\metapr} from direct fit and ratio method.}
\label{fig:dirratcomp}
\vspace{-10mm}
\end{center}
\end{figure}

\begin{figure}[t]
\begin{center}
\centerline{\epsfxsize=7.0cm 
\epsfbox{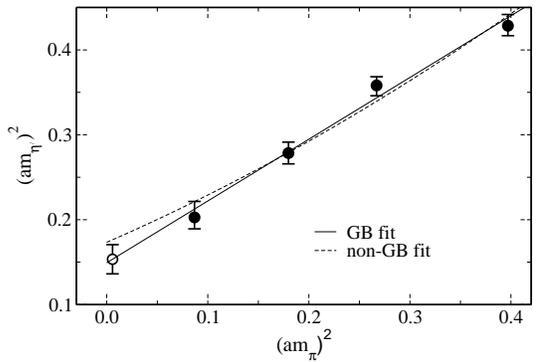} }
\vspace{-8mm}
\caption{Chiral extrapolations at \ix{\beta=2.1}.}
\label{fig:chiralxtw}
\vspace{-12mm}
\end{center}
\end{figure}

Fig.~\ref{fig:effectivemass} compares effective masses
of $\metapr$ obtained from one-quark smeared and local sources.
(Local results plotted are calculated on the 
fly by the volume source method
without gauge fixing, and have smaller errors than those
calculated from noisy sources; these have already been reported in
Ref.~\cite{cpppacslocaleta}.) 
Smearing has the clear effect of ensuring a plateau from
$t\approx 1.5$ for all parameter values.
Therefore \ix{\metapr} is determined by fitting smeared \ix{\Getapr}
for $t_{\rm min}\ge 1$ using a single hyperbolic cosine function.  We
note for this smeared data that fitting the ratio \ix{\Getapr/\Gpi}
gives a result consistent with the direct fit as shown in
Fig.~\ref{fig:dirratcomp}. The deviation remains within 8\%.




\section{$\eta^\prime$ Meson Mass in the Continuum Limit}
\label{sec:result}


We test two functional forms for chiral extrapolation,  
1) $(a\metapr)^2= A + B (a\mpi)^2$
(Goldstone boson type: `GB') and 
2) $a\metapr = A + B (a\mpi)^2$ (non-GB type) and find
that both reproduce data 
equally well (see Fig.~\ref{fig:chiralxtw} for example) 
with comparable $\chi^2/df=$ 0.9\ix{-}1.1 (GB) and 0.4\ix{-}2.2 (non-GB).
GB fit is used to determine central values, while 
non-GB fit is used for error estimation.

\begin{figure}[t]
\begin{center}
\centerline{ \epsfxsize=7.0cm 
\epsfbox{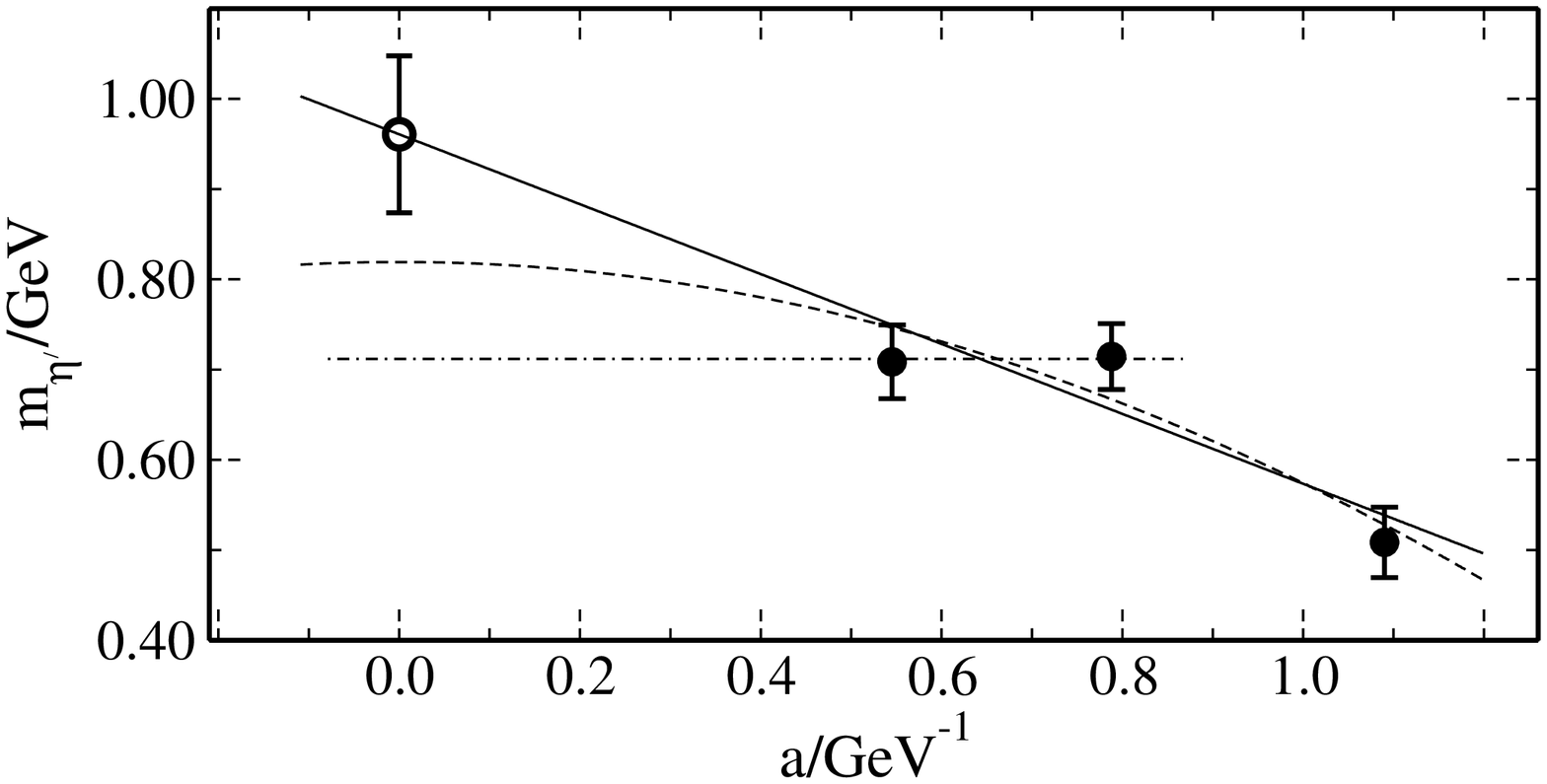}}
\vspace{-6mm}
\caption{\ix{\metapr} vs. $a$ with
linear, quadratic and constant continuum extrapolations.}
\label{fig:contx}
\vspace{-14mm}
\end{center}
\end{figure}

Fig.~\ref{fig:contx} shows $\metapr$ as a function
of $a$.
Continuum extrapolation is carried out linear in $a$,
since \ix{\Orda} error remains in our action combination.
We also try a constant plus quadratic form, since
most \ix{\Orda} effects may be
removed by use of the improved action.
Finally, since \ix{\metapr} hardly changes over the finest two
lattices, we make a constant fit removing the coarsest point.
%
The latter two are used to estimate systematic error.

Systematic errors arising from chiral and
continuum extrapolations are added in quadrature,
separately for upper and lower sides.
The final result reads\vspace{-2mm}
\begin{equation}
\metapr = 0.960(87)^{+0.036}_{-0.286} \mbox{\rm \ GeV}
\label{eq:etaresult}
\end{equation}
\vspace{-2mm}



\section{Correlation with Topological Charge}
\label{sec:topology}
\begin{figure}[t]
\begin{center}
\centerline{ \epsfxsize=7.0cm 
\epsfbox{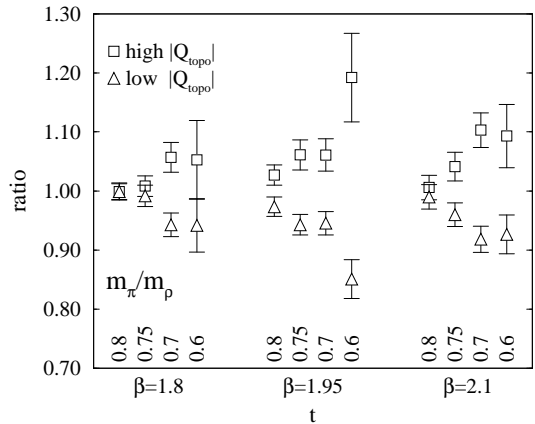}}
\vspace{-12mm}
\caption
{\ix{\metapr} from configurations with high and low \ix{|\Qtopo|}
normalized by the value for all configurations.}
\label{fig:topomsplitxx}
\vspace{-14mm}
\end{center}
\end{figure}

The large $\eta^\prime$ mass compared to the 
pion octet may originate from the topological structure of QCD.
Using results for topological charge~\cite{cppacstopology} \ix{\Qtopo},
we split the measurements of the \ix{\etapr}
propagator into two equally sized bins of high and low \ix{|\Qtopo|},
and measure \ix{\metapr} on each bin. 
As shown in Fig.~\ref{fig:topomsplitxx}, 
configurations with high \ix{|\Qtopo|} give large values of
$\metapr$ for lighter quarks,
though interpretation of results remains as open problem. 

\section{Conclusions}
Our value $\metapr$ for \ix{\Nf=2} full QCD in the continuum
limit, Eq.~\ref{eq:etaresult}, turns out to be
consistent with the experimental value, though 
there still remains considerable numerical uncertainty,
particularly on the lower side  (\ix{-30\%}),
coming from the continuum extrapolation. 
Additionally, our result contains systematic error from quenching
effects of the strange quark and from neglecting mixing with the
$\bar s s$ state.  Calculations are currently being performed
analysing mixing between quark-based states and mass eigenstates for
\ix{\Nf=2} QCD.



\vspace{2mm}
This work is supported in part by Grants-in-Aid 
of~the~Ministry~of~Education~(Nos.
P01182,   
11640294, 
12304011, 
12640253, 
13640259, 
13640260, 
14740173  
).  
VIL is a JSPS Research Fellow.





\end{document}